\documentclass[conference,9pt]{IEEEtran}
\IEEEoverridecommandlockouts
\usepackage{cite}
\usepackage{amsmath,amssymb,amsfonts}
\usepackage{graphicx}
\usepackage{textcomp}
\usepackage{multirow}
\usepackage{xcolor}
\usepackage{algorithm}
\usepackage{algpseudocode}
\def\BibTeX{{\rm B\kern-.05em{\sc i\kern-.025em b}\kern-.08em
    T\kern-.1667em\lower.7ex\hbox{E}\kern-.125emX}}
\begin{document}

\title{BitParticle: Partializing Sparse Dual-Factors to Build Quasi-Synchronizing MAC Arrays for Energy-efficient DNNs}
\author{
    Feilong Qiaoyuan$^1$,
    Jihe Wang$^1$, 
    Zhiyu Sun$^1$,
    Linying Wu$^1$,
    Yuanhua Xiao$^1$,
    and Danghui Wang$^1$
    \\$^1$School of Computer Science, Northwestern Polytechnical University, Xi'an, China
}
\maketitle
\begin{abstract}
Bit-level sparsity in quantized deep neural networks (DNNs) offers significant potential for optimizing Multiply-Accumulate (MAC) operations. However, two key challenges still limit its practical exploitation.
First, conventional bit-serial approaches cannot simultaneously leverage the sparsity of both factors, leading to a complete waste of one factor’s sparsity. Methods designed to exploit dual-factor sparsity are still in the early stages of exploration, facing the challenge of partial product explosion.
Second, the fluctuation of bit-level sparsity leads to variable cycle counts for MAC operations. Existing synchronous scheduling schemes that are suitable for dual-factor sparsity exhibit poor flexibility and still result in significant underutilization of MAC units.
To address the first challenge, this study proposes a MAC unit that leverages dual-factor sparsity through the emerging particlization-based approach. The proposed design addresses the issue of partial product explosion through simple control logic, resulting in a more area- and energy-efficient MAC unit. In addition, by discarding less significant intermediate results, the design allows for further hardware simplification at the cost of minor accuracy loss.
To address the second challenge, a quasi-synchronous scheme is introduced that adds cycle-level elasticity to the MAC array, reducing pipeline stalls and thereby improving MAC unit utilization.
Evaluation results show that the exact version of the proposed MAC array architecture achieves a 29.2\% improvement in area efficiency compared to the state-of-the-art bit-sparsity-driven architecture, while maintaining comparable energy efficiency.
The approximate variant further improves energy efficiency by 7.5\%, compared to the exact version.
\end{abstract}


\section{Introduction}
\begin{figure}[t]
\centerline{\includegraphics[width=0.5\textwidth]{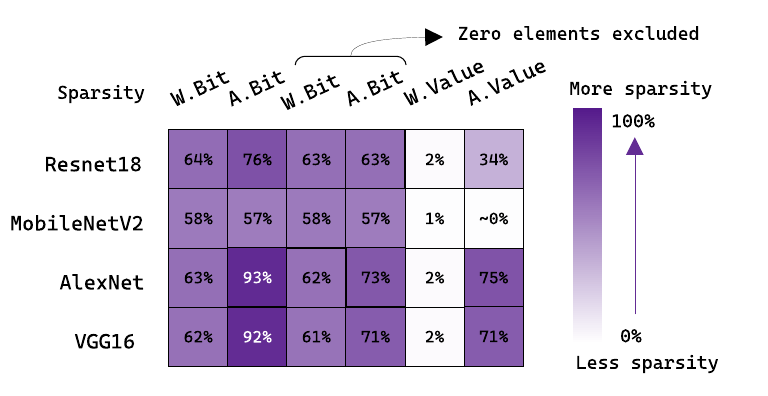}}
\caption{Sparsity of weights and activations for 8-bit quantized DNNs. Sign-magnitude representation is used instead of 2's complement, as it offers higher bit sparsity \cite{bit_sparsity_1_bitwave}.
}
\label{fig:sparsity}
\end{figure}

Due to the limited computing power of edge devices, deploying fixed-point quantized models in edge DNN architectures has become a common practice \cite{quantized_review_1,quantized_review_2}. 
However, in quantized DNNs, data sparsity still leads to resource inefficiency.
On the one hand, zero elements exist in weights and activations, known as value-level sparsity. 
Currently, numerous studies \cite{value_sparsity_1_trapezoid,value_sparsity_2_hetro_gnn,value_sparsity_3_scnn} have focused on exploiting value sparsity, that is, eliminating multiply-accumulate (MAC) operations with zero factors, to improve throughput and reduce power consumption. On the other hand, a high proportion of zero bits exists in non-zero elements, referred to as bit-level sparsity. This leads to significant useless partial product (PP) accumulations in MAC operation, since zero-valued PPs generated by these zero bits do not contribute to the result.
Previous studies \cite{bit_sparsity_1_bitwave,bit_sparsity_2_pragmatic} have shown that bit-level sparsity offers greater optimization potential than value-level sparsity, which has attracted increasing attention. The statistical data in this study (Fig. \ref{fig:sparsity}) further supports this observation.
%
%

However, existing studies on bit-level sparsity still suffer from two major limitations. First, the limited exploitation scope of sparsity. As shown in Fig. \ref{fig:sparsity}, both weights and activations exhibit bit-level sparsity potential (weights from 58\% to 63\% and activations from 57\%-71\%). However, most existing studies \cite{bit_sparsity_adas, bit_sparsity_1_bitwave,bit_sparsity_2_pragmatic,bit_sparsity_3_bitlet,bit_sparsity_4_fusekna} adopt bit-serial approaches or their variants, which can only leverage sparsity in either weights or activations, leaving the sparsity in the other operand underutilized. 
EBS \cite{bit_sparsity_ebs} introduces the idea of splitting both operands into 2-bit segments to expose dual-factor bit sparsity (e.g. 10-\textbf{00}-11-\textbf{00}). While this approach demonstrates considerable potential, it introduces a critical challenge: it results in a substantial increase in the number of partial products, thereby significantly intensifying the accumulation overhead.

Second, exploiting bit-level sparsity leads to variation in the number of cycles required for multiplication, making it difficult to fully synchronize the execution progress of the entire MAC array without incurring an unacceptable drop in MAC unit utilization. This challenge becomes even more severe under dual-factor sparsity, where both operands contain an uncertain number of zero bits, thus precluding the direct application of synchronization methods designed for single-factor sparsity. Laconic \cite{bit_sparsity_laconic} adopts a comb-style synchronization strategy that groups MAC units and performs synchronization only within each group. While this approach provides elasticity at the inter-group level, it fails to address the performance loss within a group, resulting in a portion of the benefits from sparsity inevitably being compromised.


To address the first issue, this study proposes a novel MAC unit that efficiently exploits dual-factor sparsity. Specifically, each 8-bit signed operand adopts a sign-magnitude format, which naturally offers higher bit-level sparsity compared to conventional two's complement representation \cite{bit_sparsity_1_bitwave}. Based on the concept of operand particlization, each operand is partitioned into four particles with bit-widths of 1, 2, 2, and 2, respectively. These particles are cross-multiplied to generate $4 \times 4 = 16$ products, and zero-valued products are discarded to eliminate redundant computations.
Distinct from the EBS approach, which directly accumulates these products, the proposed design treats them as intermediate results (IRs). A grouping and concatenation scheme is employed to compact the IRs into at most 7 partial products. Even in the worst-case scenario without bit-level sparsity, the number of partial products remains no greater than that of a conventional multiplier, thereby avoiding excessive accumulation overhead.

To address the second issue, this study proposes a quasi-synchronous scheduling scheme for the MAC array. In contrast to Laconic’s strict intra-group synchronization scheme, which offers no flexibility within groups, the proposed approach introduces elasticity both within and across groups. Evaluation results in Section \ref{section:4.2.3} show that even allowing intra-group flexibility alone is more effective in recovering performance loss than allowing inter-group flexibility alone. Moreover, combining both yields even better results.
Finally, at the system level, this study also introduce two switchable dataflows to accommodate convolutional (or fully connected) layers of varying shapes, providing an efficient operational scheme for the entire accelerator system.
The main contributions of this paper are as follows:

\begin{itemize}
    \item{An operand-particlization-based bit-sparsity-driven MAC unit is proposed to leverage bit-level sparsity in both weights and activations, with an approximate variant introduced to further reduces area and power consumption.}
    
    \item{A quasi-synchronous scheduling scheme is proposed to introduce cycle-level elasticity both within and across MAC unit groups, effectively mitigating utilization loss caused by pipeline stalls.}

    \item{Two switchable dataflows are proposed to implement the accelerator system, effectively accommodating convolutional (or fully connected) layers of varying shapes.}
    
    \item{Evaluation results show that the exact version of the proposed architecture achieves a 29.2\% improvement in area efficiency compared to the state-of-the-art bit-sparsity-driven architecture while maintaining comparable energy efficiency. The approximate variant further improves area efficiency and energy efficiency by 2.1\% and 7.5\%, respectively, compared to the exact version.}
\end{itemize}
\section{Background and Motivation}
\label{section:2.1}
\subsection{Limitations of Bit-serial Multiplication}
\begin{figure}[t]
\centerline{\includegraphics[width=0.5\textwidth]{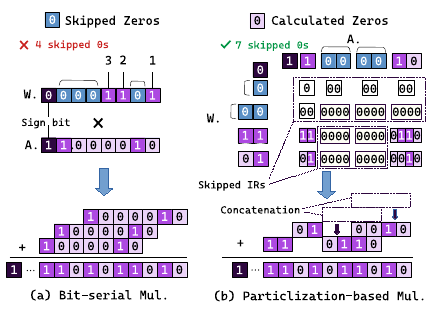}}
\caption{Bit-serial multiplication and Particlization-based multiplication.}
\label{fig:motivation}
\end{figure}
Most existing studies \cite{bit_sparsity_adas, bit_sparsity_1_bitwave,bit_sparsity_2_pragmatic,bit_sparsity_3_bitlet,bit_sparsity_4_fusekna} employ bit-serial multipliers and their variants to exploit bit-level sparsity. In this approach, the first operand is supplied to the multiplier bit by bit and multiplied with the entire second operand to generate partial products (PPs). Each PP is then shifted according to its bit position and accumulated to produce the final result. This method enables the multiplier to skip zero bits in the first operand, thereby reducing the number of PPs and lowering the overhead of PP accumulation.
For example, as illustrated in Fig. \ref{fig:motivation}(a), an ideal sparsity-driven bit-serial multiplier (it can skip all zero bits of one operand) produces three valid PPs.
However, the bit-serial multiplier fails to address a fundamental limitation: it cannot expose the bit sparsity in both operands. In the illustrated example, although the magnitude bits of two operands contain a total of 9 zeros, only 5 of them are utilized, resulting in a significant waste of sparsity potential.

\subsection{Challenges of Particlization-based Multiplication }
\label{section:2.2}
To expose bit-level sparsity from both operands, the straightforward idea is to extract all the ``1''s from both operands, pair them, and generate a number of single-bit products, which are then accumulated according to their respective bit weight to obtain the final product. However, this fine-grained approach incurs excessive hardware overhead, as it may produce up to 49 single-bit products in the worst case. To address this, EBS\cite{bit_sparsity_ebs} proposed a coarse-grained approach by partitioning operands into particles with up to 2 bits. As shown in the upper part of Fig.~\ref{fig:motivation}(b), this approach generates up to 16 products, referred to as Intermediate Results (IRs) in this study. Although this approach only exploits bit-level sparsity in ``00'' particles — discarding opportunities in ``01'' or ``10'' particles — the evaluation data presented in Section \ref{section:5.3} shows that its utilization of bit-level sparsity still surpasses that of the ideal bit-serial approach, demonstrating the promise of this idea.

However, EBS still faces a critical issue: the worst-case scenario of 16 IRs is still significantly larger than the number of PPs in a conventional multiplier (at most $7$ for 8-bit sign-magnitude operand). The shift-and-accumulate process for these IRs incurs considerable energy overhead and degrades throughput, which offsets the benefits of exploiting bit-level sparsity. Furthermore, the EBS design breaks the traditional MAC unit boundary by mixing particles from different MAC operations. As a result, it must maintain additional metadata for each particle, including its signed bit and least significant bit (LSB) weight, which further increases hardware overhead. In summary, the particlization-based approach has not been thoroughly investigated, and its design details call for further optimization.

To address the above issues, this study proposes a novel MAC unit that adopts particlization-based multiplication with several key improvements. The most critical improvement lies in the observation that many IRs do not overlap in their bit positions, allowing a significant portion of the accumulation to be replaced by zero-overhead concatenation (as shown in the lower part of Fig.\ref{fig:motivation}(b)). Through appropriate grouping and concatenation, the number of IRs can be reduced to at most 7, which matches the PP count of a conventional multiplier and significantly reduces accumulation overhead.
Second, during the selection of nonzero IRs, each multiplexer is responsible for IRs with a fixed LSB weight. This not only eliminates the need to explicitly maintain weight metadata but also ensures a fixed shift distance, allowing barrel shifters to be replaced with zero-overhead wire shifting. Finally, the proposed design maintains the independence of the MAC unit, so only one sign bit per operand needs to be maintained, rather than one per particle.

\subsection{Synchronization Issue in the MAC Array}
\label{section:2.3}
\begin{figure}[t]
\centerline{\includegraphics[width=0.5\textwidth]{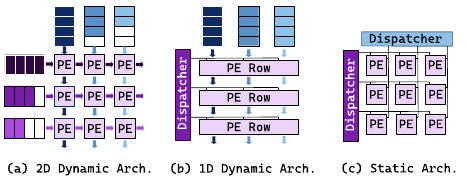}}
\caption{Dynamic and static architectures.}
\label{fig:dynamic_static_arch}
\vspace{-1em}
\end{figure}

Building a MAC array with the bit-sparsity-driven MAC units presents another challenge. Traditional MAC units have a fixed number of cycles to perform a MAC operation, whereas bit-sparsity-driven MAC units usually have an uncertain number of cycles due to fluctuations in the number of zero bits, which leads to additional challenges.
To clarify this issue, MAC Array architectures are classified into two categories based on whether data propagates across PEs (Processing Elements, where each PE is equivalent to a MAC unit in this study): dynamic and static architectures. As shown in Fig. \ref{fig:dynamic_static_arch}, a Systolic Array \cite{dataflow_systolic}  represents a 2-dimensional dynamic architecture, where weights and activations enter the PEs from two directions and propagate along both row and column after each computation (Fig. \ref{fig:dynamic_static_arch}(a)). Trapezoid \cite{value_sparsity_1_trapezoid}, on the other hand, is a 1D dynamic architecture where data along one dimension is statically loaded into PEs, while data along the other dimension propagates across PEs (Fig. \ref{fig:dynamic_static_arch}(b)). In dynamic architectures, the use of PEs with variable computation cycles without proper synchronization can cause misaligned computation, ultimately leading to incorrect results.

 In contrast, static architectures, such as BitWave\cite{bit_sparsity_1_bitwave}, employ data distribution logic to directly assign operands to each PE, eliminating the need for operand propagation between PEs, as illustrated in Fig.~\ref{fig:dynamic_static_arch}(c). However, even in static architectures, computation misalignment can lead to irregular memory access patterns. Since a single cache bank may serve multiple PEs within one cycle, variations in PE progress may require multiple cache lines to be accessed within a single cycle, which presents a challenge to memory bandwidth and access regularity. In summary, to prevent computation errors and mitigate memory access irregularities, appropriate mechanisms must be introduced to synchronize PE progress. These mechanisms, however, inevitably introduce pipeline bubbles, which degrade PE utilization.

In this study, since both weights and activations exhibit fluctuating bit-level sparsity, a synchronization scheme is required that can adapt to dual-factor sparsity. Among existing works, Laconic\cite{bit_sparsity_laconic} proposes a comb-style synchronization scheme, which partitions PEs into groups. While this scheme permits progress divergence across groups, it enforces strict cycle-level synchronization within each group, leaving no room for cycle elasticity. As a result, the PE utilization loss caused by intra-group synchronization remains unresolved. 
To address this limitation, this study proposes a quasi-synchronous scheduling scheme. In this scheme, each column of the 2D MAC array is treated as a group, and cycle elasticity is introduced both within and across groups using a simple buffering mechanism.
Evaluation results demonstrate that allowing intra-group cycle elasticity is more effective in mitigating PE utilization loss compared to inter-group cycle elasticity. In addition, this study proposes two dataflows designed to align with the proposed quasi-synchronization scheme for constructing the accelerator system. Each dataflow is optimized for convolution (or fully connected) layers of different shapes, mitigating MAC unit idleness caused by insufficient parallelism.

\section{Sparsity-driven MAC Unit}
\label{section:3}
\subsection{Methodology}
\label{section:3.1}
\begin{figure}[t]
\centerline{\includegraphics[width=0.5\textwidth]{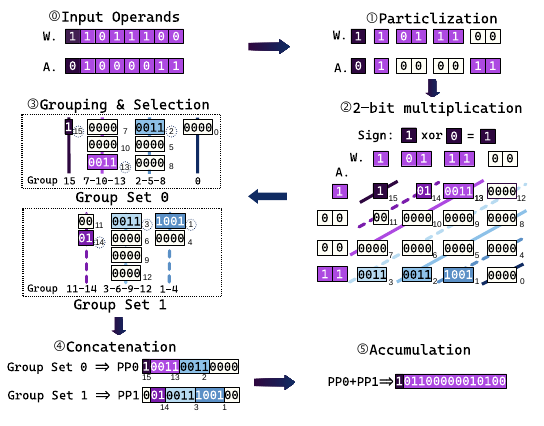}}
\caption{Five steps for leveraging dual-factor sparsity.}
\label{fig:particlization}
\end{figure}

Fig. \ref{fig:particlization} illustrates the computation flow of the proposed MAC unit. The process begins with the sign bits: the sign bits of the two operands are extracted and XORed to determine the sign bit of the final product. The remaining magnitude parts are then treated as a 7-bit unsigned multiplication. The handling of the magnitude part involves five steps.
First, the 7 magnitude bits are divided into particles of 1, 2, 2, and 2 bits.
Second, particles from the two operands are paired and multiplied, generating 16 intermediate results (IRs) arranged into a $4 \times 4$ IR matrix. Each IR is assigned a unique position ID from $0$ to $15$.
Each IR has a bit width of at most $4$, and its LSB weight is solely determined by its position in the matrix. This property ensures that certain IRs will never overlap with others. Based on this observation, the third step is to group IRs with the same LSB weight, which lie along lines parallel to the matrix’s anti-diagonal (as indicated by colored lines in the figure). Each group is named after the IDs of its IRs, such as group 3-6-9-12.

The seven groups are divided into two sets. Group Set 0 includes group 15, group 7-10-13, group 2-5-8, and group 0, while Group Set 1 contains the remaining three groups. Since IRs in different groups within a group set do not overlap, selecting one IR from each group allows them to be concatenated into a single value of up to 13 bits. By analogy with traditional multipliers, this value is referred to as a partial product. Consequently, even in the worst case, only seven PPs are generated: group 3-6-9-12 in Group Set 1 can be selected at most four times, and group 7-10-13 or group 2-5-8 in Group Set 0 can be selected at most three times.
Finally, these PPs are all accumulated to obtain the final product. Through grouping and concatenation, the 16 IRs are reduced to no more than $7$ PPs—matching the number generated in a conventional $7$-bit multiplier—thereby effectively controlling the growth in PPs. Efficiently selecting nonzero IRs to be concatenated is key to exploiting bit-level sparsity, and the detailed microarchitecture will be described in the next subsection.

\subsection{Microarchitecture}
\label{section:3.2}
\begin{figure}[t]
\centerline{\includegraphics[width=0.5\textwidth]{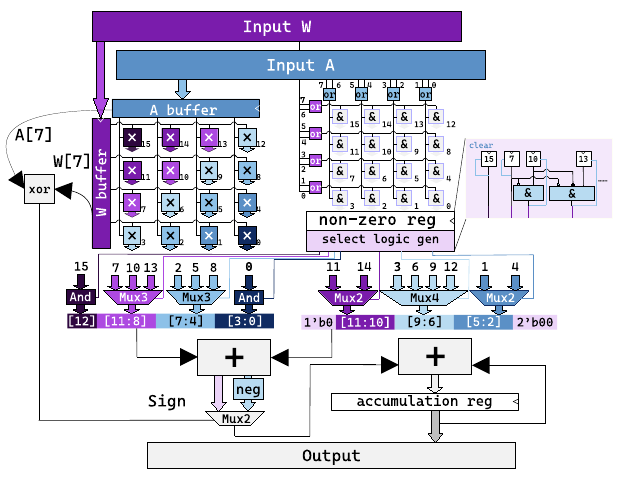}}
\caption{Proposed sparsity-driven MAC unit}
\label{fig:mac_unit}
\end{figure}
\subsubsection{Data Path and Cycle Scheduling}
Fig. \ref{fig:mac_unit} shows the microarchitecture of the proposed MAC unit. 
During the first cycle,  the input operands $W$ and $A$ are each written into a buffer (shown in the upper left part of the figure).
Then, during the following 1 to 4 cycles, two PPs are generated and accumulated in each cycle. Data read from the buffer first passes through a matrix of 2-bit multipliers to produce 16 IRs. According to the grouping rule, IRs within the same group are connected to a one-hot multiplexer (It is worth noting that one-hot multiplexers have simpler logic compared to normal multiplexers, as they directly accept ``decoded" control signals and do not require internal decoding logic). 
In each cycle, one non-zero IR is selected from each group, and two PPs (from Group Set 0 and Group set 1) are formed through static concatenation. Generated two PPs are added by a 13-bit adder, then converted from sign-magnitude to 2’s complement format (for easier accumulation with opposite signs), and accumulated to the accumulation register.
Finally, once all groups have at most one non-zero IR remaining in the current cycle, indicating that no more IRs are available for selection in the next cycle, the computation completes, and new operands can be written into the buffer in the same cycle. Since the last cycle of the current MAC operation overlaps with the first cycle of the next MAC operation, new MAC operations can be accepted with an initiation interval of 1 to 4 cycles.

\subsubsection{Control Logic}
The control logic is shown in the upper right part of the figure. In the first cycle, an OR operation within each particle detects non-zero particles. These results are processed through a cross-AND array to generate a 16-bit non-zero vector, which indicates whether each IR is non-zero and written into the non-zero register. In the following 1-4 cycles, each group independently generates one-hot encoded control signals through priority selection logic. If an IR is selected, the corresponding bit in the non-zero register is cleared in the next cycle to prevent reselection. 

\subsubsection{Optimizing 2-bit Multiplication}
An optimization technique is introduced for implementation: a 4-bit IR can actually be represented using 3 bits via encoding $9$ as $111$, as it has only seven possible values: $0$, $1$, $2$, $3$, $4$, $6$, and $9$. The circuit required to generate a 3-bit IR is much simpler than that for 4 bits, and importantly, this reduces the bit width of the one-hot multiplexers, which are a primary cost in the control logic. The 3-bit IRs need to be converted back to 4 bits before accumulation, but this conversion circuit only requires 5 instances. With this optimization, the area and power of the MAC unit are each reduced by 2–3\%.

\subsubsection{Approximating for Area and Power Reduction}
Multiplying two 8-bit sign-magnitude formatted numbers produces a 15-bit result, but the input data for the next layer is quantized back to 8 bits, meaning that several of the least significant bits are ultimately discarded. This observation suggests that IRs associated with smaller LSB weights can be omitted without significantly affecting the final output accuracy.
This study proposes an approximate variant of the MAC unit, which unconditionally discards all IRs from groups 1–4 and group 0, regardless of their value.
This optimization reduces the area and power overhead associated with computing and accumulating these IRs. Implementation is straightforward — the corresponding IR computation and selection circuits are simply removed, along with the associated control logic.

The proposed approximate variant was evaluated using the ResNet-18 model \cite{eval_model_resnet} on the CIFAR-10 dataset \cite{eval_dataset_cifar10}. Both weights and activations were quantized to 8-bit using per-tensor symmetric quantization. All MAC operations in convolutional and fully connected layers were replaced with the proposed approximate version. Experimental results show that the model achieves an inference accuracy of 90.2\%, compared to 93.8\% for the exact version. This results in only a 3.6\% accuracy drop, while enabling a 20.0\% reduction in area and a 13.6\%–15.1\% reduction in power consumption for a MAC unit under bit sparsity levels ranging from 50\% to 90\%, offering designers a compelling trade-off between accuracy and hardware cost.
Additionally, the reduced IR count slightly decreases the average number of cycles required per multiplication. However, since the two discarded groups rarely become the bottleneck, the reduction in average cycles is less than 1\%. Therefore, the primary benefit of this optimization stems from the savings in area and power consumption.

\section{Dataflow and Synchronization}
To provide a clearer understanding of the context in which the synchronization issue arises, this section first introduces the dataflow in Subsection A, and then presents the quasi-synchronization scheme in Subsection B.
\subsection{Two Types of Dataflows Adapt to Differently Shaped Layers}
\label{section:4.1}
\subsubsection{Introduction to Dataflow}
\begin{figure}[t]
\centerline{\includegraphics[width=0.5\textwidth]{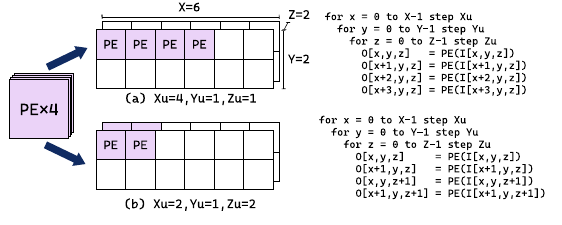}}
\caption{An example of dataflow}
\label{fig:dataflow}
\end{figure}

With a fixed number of PEs and variable workload across NN layers, each PE must handle multiple MAC operations, creating a task allocation challenge. This allocation, or \emph{dataflow}, determines how tasks are assigned to PEs.
For example, as shown in Fig.~\ref{fig:dataflow}, a triply nested for loop spans the X, Y, and Z dimensions, forming a three-dimensional iteration space.
In dataflow (a), four PEs operate in parallel along the X dimension, while the Y and Z dimensions are processed sequentially. Each iteration of the innermost loop covers a $4 \times 1 \times 1$ subspace.
Following the notation and concepts introduced in prior research \cite{bit_sparsity_1_bitwave}, this configuration is described as spatial unrolling along the X dimension with a parallelism factor of $X_u=4$, and temporal unrolling along the Y and Z dimensions (with $Y_u=1$ and $Z_u=1$).
However, this dataflow results in low PE utilization, as each cycle of the second iteration in the X dimension has only two valid computations. In contrast, dataflow (b) with $X_{u}=2$, $Y_{u}=1$, and $Z_{u}=2$ avoids the inefficiency. Overall, designing a dataflow involves mapping PEs across different dimensions to form a fixed-volume hypercube, which iterates repeatedly until the entire high-dimensional iteration space is covered.

\begin{table}[t]
    \centering
    \scriptsize
    \caption{$7$ dimensions of DNN computation for convolutional layer}
    \begin{tabular}{|c|c|c|c|c|}
        \hline
        Dimension & Description & W. Reuse & A. Reuse\\
        \hline
        B & Batch & True & Flase\\
        K & Output Channel/Kernel& False & True \\
        C & Input Channel & False & False \\
        OY & Output Height & True & True\\
        OX & Output Width & True & True\\
        FY & Kernel Height & False & False \\
        FX & Kernel Width & False & False \\
        \hline
    \end{tabular}
    \label{tab:dnn_dim}
\end{table}

TABLE \ref{tab:dnn_dim} summarizes the seven dimensions of a convolutional layer. It is observed \cite{bit_sparsity_1_bitwave} that the early layers of DNNs tend to have larger OX and OY dimensions and smaller C and K dimensions, while the later layers exhibit the opposite trend. This variation in layer shapes necessitates adaptable dataflows to maximize PE utilization. Furthermore, since weights and activations can be reused across certain dimensions (e.g., different batches process distinct activations but share the same weights), the dataflow plays a crucial role in determining memory access pattern.

\subsubsection{Proposed Dataflow}
\begin{figure}[t]
\centerline{\includegraphics[width=0.5\textwidth]{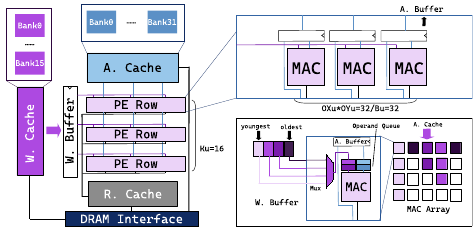}}
\caption{MAC array architecture. The left and upper-right parts of the figure illustrate the accelerator system architecture and the basic design of the MAC row, respectively. The lower-right part shows the additional logic introduced to implement the quasi-synchronization scheme.}
\label{fig:mac_array}
\end{figure}

In this subsection, a MAC unit is assumed to perform one MAC operation per cycle for simplified understanding, with further details addressed later.
As shown in the left part of Fig. \ref{fig:mac_array}, the proposed architecture includes a 2D PE array with 16 rows and 32 columns.  Weights, activations, and temporary accumulated results are stored in separate caches that exchange data with off-chip DRAM. The weight cache has 16 banks, each corresponding to a PE row, while the activation cache and result cache have 32 banks, each corresponding to a PE column. In each cycle, 16 weights are loaded into the weight buffer and each is shared across a row, while 32 activations enter from the top of the PE array, propagating down through the rows until they exit.

The MAC array requires reusing weights across different columns and activations across different rows, enabling two types of dataflows: $K_{u}=16$ along columns, while (a) $OX_{u} \times OY_{u}=32$ or (b) $B_{u}=32$ along rows. Dataflow (a) fits early convolutional layers with large $OX$ and $OY$, while dataflow (b) fits later convolutional or fully connected layers. Spatial unrolling along both $OX$ and $OY$ enhances PE utilization when $OX < 32$. The values of $OX_u$ and $OY_u$ are configurable, with three possible combinations: \( (OX_u, OY_u) \in \{(32, 1), (16, 2), (8, 4)\} \). The proposed dataflows have two advantages: (a) Reuse of row and column data within the PE array, reducing memory bandwidth demand; (b) The four dimensions involved in spatial unrolling—$K$, $B$, $OX$, and $OY$— produce independent results, which means that the outputs generated by different PEs do not require accumulation, eliminating the need for inter-PE accumulators.

In this design, the weight, activation, and result caches are configured to 64KB, 128KB, and 128KB, respectively. However, design is not limited to the above cache configurations. Under stricter area constraints, smaller caches can still maintain a high hit rate by leveraging loop tiling \cite{system_loop_tiling}, a widely adopted optimization technique, which splits large data dimensions into smaller tiles to better fit the on-chip storage.
For instance, in dataflow (a), computations can be scheduled as \(B\)-\(OY_1\)-\(OX_1\)-\(K_1\)-\(FY\)-\(FX\)-\(C\).
The tiling of \(K\), \(OX\), and \(OY\) is required because \(K\) must be spatially unrolled with a parallelism factor of 16, leading to \(K\) being split into two nested loops, \(K_0\) and \(K_1\), with \(K_0 = 16\) and \(K = K_0 \times K_1\). \(OX\) and \(OY\) are similarly partitioned for spatial unrolling. The weight cache stores all the data required for the innermost \(K_1\)-\(FY\)-\(FX\)-\(C\) loops, enabling reuse at \(OX_1\).
If the cache capacity is insufficient, a scheduling strategy such as \(B\)-\(C_1\)-\(OY_1\)-\(OX_1\)-\(K_1\)-\(FY\)-\(FX\)-\(C_0\) can be adopted, where \(C\) is partitioned to reduce the data volume of the lowest four dimensions, effectively addressing the issue.

To enable efficient execution of a model on the proposed architecture, a mapping strategy — including the dataflow and loop tiling — must be determined for each NN layer under a given cache configuration. This is a well-established problem, and existing tools such as ZigZag\cite{eval_zigzag} are able to automatically generate an optimal or near-optimal mapping once the NN model, memory hierarchy, and candidate dataflows are specified.

\subsection{Quasi-Synchronizing Scheme}
\label{section:4.2}
The last subsection assumes the MAC unit to be single-cycled. However, when using bit-sparsity-driven MAC units, any MAC unit within the 2D array could take an uncertain number of cycles, leading to synchronization issues, resulting in reduced PE utilization and complex memory access patterns.  
To address this issue, the proposed quasi-synchronization scheme groups the MAC units, with each column of MAC units treated as a group, resulting in a total of 32 groups. Elasticity is introduced in two dimensions: intra-group and inter-group.

\subsubsection{Intra-group elasticity via buffering queues} 
As shown in the bottom-right part of Fig.~\ref{fig:mac_array}, the proposed scheme requires all MAC units within a group to propagate one step forward synchronously, meaning no additional bubbles are allowed to be inserted. However, enforcing strict synchronization alone would force all MAC units in the group to wait for the slowest one to complete its computation before propagating data to the next PE.
To mitigate this constraint, each MAC unit is equipped with an additional operand queue with a capacity of $Q=2$, allowing it to buffer the operands of two incoming MAC operations. Once a MAC operation is enqueued, it is considered accepted. If all operations in the group are accepted within the current cycle, the group is allowed to propagate to the next step. Within each MAC unit, variations in computation latency across consecutive operations are smoothed out by the queue.
As $Q$ increases, the throughput of each MAC unit gradually approaches its ideal standalone performance, becoming less affected by the latency of other units in the same group. However, a larger $Q$ also implies increased hardware overhead, thus requiring a trade-off.

Introducing the buffering queue also brings an additional key benefit: zero operands can be pre-filtered and prevented from entering the queue (referred to as zero-value filtering), thereby reducing queue pressure. As a result, each MAC unit only needs to process non-zero multiplications, reducing the actual cycle cost of a zero-valued multiplication from 1 to 0. This optimization can significantly improve the throughput of the MAC array in scenarios with high value sparsity.

\subsubsection{Inter-group elasticity via weight selection} 
In addition to using buffering queues to add elasticity within a group, the proposed scheme allows a step divergence of up to $E=3$ across groups. Specifically, the fastest group is allowed to advance up to $E$ steps ahead of the slowest one. This mechanism enables faster groups to proceed without waiting for slower ones, thereby reducing idle time for the MAC units.
However, since weights are shared among MAC units located in the same row, which belong to different groups, step divergence introduces a consistency challenge. Slower groups will require ``older'' weights, while faster groups will require ``younger'' ones. This implies that each weight must be retained in a buffer until it has been consumed by the slowest group.
As illustrated in the bottom-right part of Fig.~\ref{fig:mac_array}, the two darkest-colored groups represent the slowest progress, lagging $E=3$ steps behind the fastest (lightest-colored) group. To ensure correctness, the system must store the latest $E+1=4$ weights in the weight buffer and equip each MAC unit with a multiplexer to select the appropriate weight.
\subsubsection{Improvements in PE Utilization and Average Cycle per Step}
\label{section:4.2.3}
To demonstrate the effectiveness of the proposed quasi-synchronization scheme and to provide guidance for selecting $E$ and $Q$, this study conducts a series of ablation experiments.
A cycle-accurate simulator is developed for the MAC array, which accepts weights and activations randomly generated following a defined distribution. To eliminate the influence of unrelated factors, the simulator ensures that as long as a column is ready to advance, sufficient input data is always available, preventing stalls or idling caused by cache misses or limited spatial parallelism.

\begin{figure}[t]
\centerline{\includegraphics[width=0.48\textwidth]{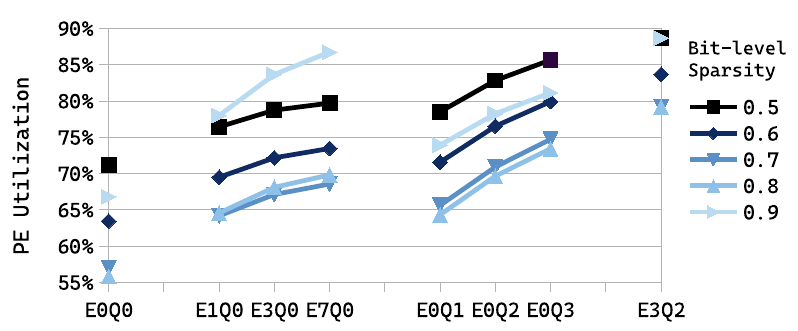}}
\caption{PE utilization under different E,Q and bit-level sparsity.(range form 55\% to 90\%)}
\label{fig:eval_util}
\end{figure}

\begin{figure}[t]
\centerline{\includegraphics[width=0.48\textwidth]{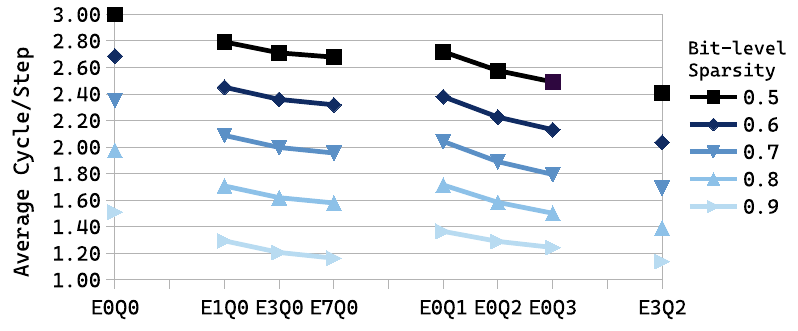}}
\caption{Average number of cycles required for data to propagate one step forward under different E,Q and bit-level sparsity (lower is better).}
\label{fig:eval_cycle}
\end{figure}

The first set of experiments investigates how different levels of intra-group and inter-group elasticity affect PE utilization and performance. Each configuration is denoted as $ExQy$, where $E = x$ and $Q = y$. For example, $E0Q0$ indicates that neither inter-group nor intra-group elasticity is enabled. The values of $E$ are chosen as $2^k - 1 (k \in \mathbb{N})$ to reduce address space waste in the weight buffer.
Five subsets of experiments are conducted using datasets with bit sparsity ($bs$) of 0.5, 0.6, 0.7, 0.8, and 0.9. The data generation process assigns each bit a probability of $bs$ to be 0 and $(1-bs)$ to be 1 independently. 
In this set of experiments, zero-value filtering is disabled, meaning zero-valued operands are still inserted into the queue and processed as regular computations.

Fig.~\ref{fig:eval_util} shows how PE utilization varies across different configurations and bit sparsity levels. The results can be summarized as follows:
(1) Introducing either inter-group elasticity ($ExQ0$) or intra-group elasticity ($E0Qy$) leads to a substantial improvement in PE utilization compared to the baseline (E0Q0), and combining both boosts utilization from 55.8\%-71.2\% to 79.1\%-88.7\%.
(2) The improvements brought by increasing either intra-group or inter-group elasticity exhibit diminishing returns. For example, when the bit sparsity is 0.7, increasing $E$ from 1 to 3 ($E1Q0$ → $E3Q0$) improves PE utilization by 2.9\%, while further increasing $E$ to 7 yields only a 1.4\% gain, at the cost of doubling the required buffer capacity and increasing the complexity of the multiplexer for weight selection. Considering both hardware cost and performance benefit, $E3Q2$ represents a relatively balanced configuration.  

(3) For bit sparsity levels between 0.5 and 0.8 — a range representative of most DNN workloads — intra-group elasticity consistently outperforms inter-group elasticity, underscoring its practical effectiveness. The noticeable performance gain of inter-group elasticity at a sparsity of 0.9 can be attributed to the fact that, under such extreme sparsity, most computations are completed within a single cycle. When a group stalls, other groups typically have few remaining computations and quickly enter an idle state. As a result, the utilization loss caused by stalls becomes more pronounced at extremely high sparsity, which magnifies the advantage of inter-group elasticity.
Fig.~\ref{fig:eval_cycle} presents the average number of cycles required for each group to advance one step, providing a more intuitive view of the performance improvements and validating the three conclusions.



\begin{figure}[t]
\centerline{\includegraphics[width=0.52\textwidth]{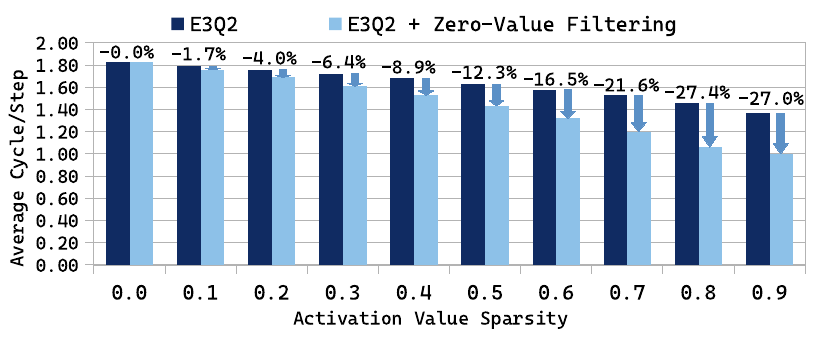}}
\caption{Impact of zero-value filtering on average cycle per step (lower is better; configuration:weight 
 value sparsity = 0; weight/activation bit sparsity = 0.65)}
\label{fig:eval_cycle_valuesparsity}
\end{figure}

The second set of experiments evaluates the effectiveness of zero-value filtering, with performance measured by the average cycles per step. The configuration is fixed to $E3Q2$, while the bit-level sparsity of non-zero elements is set to 0.65 — a typical value in DNNs — and the weight value sparsity is ignored, as it is usually below 0.1.  
As shown in Fig.~\ref{fig:eval_cycle_valuesparsity}, the effectiveness of zero-value filtering increases as the activation value sparsity rises. When the activation value sparsity reaches 0.8, zero-value filtering reduces the average cycles per step by 27.4\%, corresponding to a throughput (average steps per cycle) improvement of 37.7\%.

Furthermore, the evaluation is extended using data generated based on the statistical sparsity patterns of real DNN models. Under these conditions, zero-value filtering improves the MAC array throughput by 7.9\%, 0.1\%, 30.4\%, and 28.8\% for ResNet18\cite{eval_model_resnet}, MobileNetV2\cite{eval_model_mobilenetv2}, AlexNet\cite{eval_model_alexnet}, and VGG16\cite{eval_model_VGG}, respectively. MobileNetV2 shows minimal improvement due to its extremely low value sparsity, while the other networks benefit significantly.

\section{Evaluation}
\label{section:5}
\subsection{Methodology}
\label{section:5.1}
\subsubsection{Modeling}
The proposed MAC unit is modeled in Verilog HDL at the register-transfer level (RTL), and its area and power consumption are analyzed using Synopsys Design Compiler (DC) under a 45nm process technology at 500 MHz.
For the MAC array evaluation:
(1) Operand queues and control logic introduced by the quasi-synchronizing approach are integrated into the MAC unit’s RTL model, and both area and average energy per operation are estimated using DC.
(2) Memory components, including other buffers, on-chip SRAM, and off-chip DRAM, are modeled using CACTI-IO \cite{eval_cacti} to estimate their access energy, and the area occupied by all on-chip buffers and SRAMs is also evaluated.
(3) The cycle-accurate MAC array simulator introduced in section \ref{section:4.2.3} measures the average cycles per MAC operation using data generated based on model-specific distributions.
(4) The results from (1)-(3) are used to calculate the total chip area. In addition, the ZigZag\cite{eval_zigzag} accelerator modeling tool is employed to model the accelerator system, with per-access energy for each memory level, the number of MAC units, the average cycles per MAC operation, and the energy per MAC operation provided as configuration parameters.
(5) ZigZag selects a specific dataflow and memory mapping strategy for each NN layer, and simulates execution under realistic conditions, including cache misses and insufficient parallelism. It reports actual memory access and MAC operation counts, enabling accurate system-level power estimation.
All evaluations are based on the same 45nm process to ensure consistency.

\subsubsection{Experimental Setup and Configurations}
The experiments are organized into three sections. Section \ref{section:5.2} analyzes the performance, area, and power consumption of the proposed MAC unit, comparing it against the state-of-the-art sparsity-driven MAC units. Section \ref{section:5.3} introduces the enhancement in sparsity exploitation achieved by the proposed approach over the bit-serial approach. Section \ref{section:5.4} evaluates the MAC array architecture in comparison to state-of-the-art approaches.
The comparison is made with two state-of-the-art bit-sparsity-driven accelerators, BitWave (2024) \cite{bit_sparsity_1_bitwave} and AdaS (2023) \cite{bit_sparsity_adas}. These accelerators are also modeled using ZigZag under a 45nm process, with the basic metrics for the PEs extracted from the RTL model implemented in Verilog HDL. The number of PEs and the cache size are listed in TABLE \ref{tab:configurations}.

\begin{table}[t]
    \centering
    \caption{Configurations of DNN Accelerators}
    \scriptsize
    \begin{tabular}{|c|c|c|c|}
       \hline
        \textbf{Configuration} & BitParticle & BitWave(2024)\cite{bit_sparsity_1_bitwave} & AdaS(2023) \cite{bit_sparsity_adas} \\
        \hline
        PE count & 512 & 512 (BCEs) & 64 (\(\times\)4 Lanes) \\
        W. Cache & 64KB & 256KB & 128KB \\
        A. Cache & 128KB & 256KB & 128KB \\
        R. Cache & 128KB & shared with A. & / \\
        MetaData Buf. & / & / & 64KB \\
        \hline
    \end{tabular}
    \label{tab:configurations}
\end{table}

\subsubsection{Networks and Datasets}
The experiment uses four DNN models: ResNet18\cite{eval_model_resnet}, MobileNetV2\cite{eval_model_mobilenetv2}, VGG16\cite{eval_model_VGG}, and AlexNet\cite{eval_model_alexnet}, with respective number of parameters of 11.7M, 3.5M, 61.1M, and 138.4M. These models span a range of scales and are evaluated on the CIFAR-10\cite{eval_dataset_cifar10} and STL-10\cite{eval_dataset_stl10} datasets. The CIFAR-10 and STL-10 datasets both consist of real-world RGB images across ten object categories, with resolution of $32 \times 32$ and $96 \times 96$, respectively.

\subsection{Efficiency-Enhanced Sparsity-Driven MAC Unit}
\label{section:5.2}
\begin{table}[t]
\centering
\caption{Performance, power, and area statistics for different types of MAC units}
\resizebox{0.48\textwidth}{!}{
\begin{tabular}{|c|c|c|c|c|c|c|c|}
\hline
& Sparsity & AdaS\cite{bit_sparsity_adas} & BitWave\cite{bit_sparsity_1_bitwave} & BP-exact & BP-approx\\
\hline
\multirow{5}{*}{\textbf{Average Cycles/OP}} 
& 50\%  & 3.22 & 0.91 & 2.14 & 2.12 \\
& 60\%  & 2.46 & 0.85 & 1.71 & 1.69 \\
& 70\%  & 1.80 & 0.76 & 1.34 & 1.33 \\
& 80\%  & 1.29 & 0.62 & 1.10 & 1.10 \\
& 90\%  & 1.04 & 0.42 & 1.01 & 1.01 \\
\hline
\textbf{Area ($\mu$m\textsuperscript{2})} & / & 462.04 & 1504.76 & 544.50 & 443.42  \\
\hline
\multirow{5}{*}{\textbf{Power ($\mu$W)}} 
& 50\% & 439.81 & 1054.50 & 509.38 & 432.20  \\
& 60\% & 434.80 & 1008.10 & 481.01 & 409.94 \\
& 70\% & 420.49 & 923.44 & 451.49 & 386.40  \\
& 80\% & 368.47 & 867.41 & 392.54 & 339.17 \\
& 90\% & 285.83 & 728.43 & 318.13 & 273.24 \\
\hline
\multirow{4}{*}{\textbf{Normalized}} 
& 50\% & 1.00 & 1.08 & \textbf{1.28} & 1.58 \\
\multirow{4}{*}{\textbf{Area Efficiency}} 
& 60\% & 1.00 & 0.89 & \textbf{1.23} & 1.52 \\
& 70\% & 1.00 & 0.72 & \textbf{1.14} & 1.41 \\
& 80\% & \textbf{1.00} & 0.64 & 0.99 & 1.23 \\
& 90\% & \textbf{1.00} & 0.76 & 0.87 & 1.07 \\
\hline
\multirow{4}{*}{\textbf{Normalized}} 
& 50\% & 1.00 & \textbf{1.47} & 1.30 & 1.55\\
\multirow{4}{*}{\textbf{Energy Efficiency}} 
& 60\% & 1.00 & 1.24 & \textbf{1.31} & 1.55\\
& 70\% & 1.00 & 1.07 & \textbf{1.25} & 1.47\\
& 80\% & 1.00 & 0.88 & \textbf{1.10} & 1.28\\
& 90\% & \textbf{1.00} & 0.97 & 0.92 & 1.07\\
\hline
\end{tabular}}
\label{tab:performance_comparison}
\end{table}

This section demonstrates the advantages of the proposed MAC unit (BitParticle), over state-of-the-art sparsity-driven approaches by analyzing its area, power, average cycles per operation, energy efficiency (TOPS/W), and area efficiency (TOPS/mm$^2$) when operating standalone. Both area efficiency and energy efficiency are normalized to AdaS.
The evaluated MAC units include the exact (BP-exact) and approximate (BP-approx) versions of BitParticle, and two prior works: AdaS and BitWave. For fair comparison, each MAC unit is defined as the minimal logic required to independently complete MAC operations. Specifically, BitWave requires both BCE and Zero-Column Index Parser to complete a MAC, so both are included. Since BitWave handles 8 MACs per computation round, its average cycles per operation are computed as total cycles divided by 8. For AdaS, the unrelated Inner-Join module is excluded, and a single-lane version (one MAC per round) is evaluated.
Input data is generated as described in Section~\ref{section:4.2.3}, covering bit-sparsity levels from 0.5 to 0.9.

TABLE \ref{tab:performance_comparison} shows the experimental results; the bolded data indicates the best approach (counting only the exact version). Compared to AdaS, BP-exact demonstrates superior area efficiency across data with 50\%-70\% bit sparsity and is comparable at 80\%. For typical bit sparsity levels of neural networks (60\%-70\%), BP-exact provides improvements of 23\%-14\% in area efficiency and 31\%-25\% in energy efficiency. 
The key reason for BitParticle's advantage is that it achieves the accumulation of two partial products per cycle at a minimal 
 additional area and power cost, owing to the simplicity of the control logic for selecting and assembling IRs. 
 
 Compared to BitWave, BP-exact improves energy efficiency by 5.6\%, 16.8\%, and 25.0\% at 60\%, 70\%, and 80\% bit sparsity, respectively. The main drawback of the BitWave is that it processes eight MAC operations in parallel and can skip computations only when the same bit position is zero across all eight data elements, which results in limited exploitation of bit-level sparsity. Moreover, BitWave suffers from a significant drawback in the area; BP-exact achieves higher area efficiency across data with 50\%-90\% bit sparsity, with improvements of 38.2\%-58.3\% at 60\%-70\% sparsity levels.

Finally, the approximate version of BitParticle demonstrates significant advantages by reducing hardware complexity, achieving approximately 23\% and 18\% improvements in area efficiency and energy efficiency, respectively, compared to the exact version.

\subsection{Superior Sparsity Utilization Compared to Bit-Serial}

This section compares the particlization-based approach with the bit-serial approach to quantify how much additional sparsity potential can be exploited by leveraging dual-factor sparsity, which was not discussed in EBS\cite{bit_sparsity_ebs}. First, a metric called \emph{Skipped Calculations} is defined to assess how different approaches leverage bit-level sparsity.
In the sign-magnitude representation, an 8-bit number consists of a sign bit and 7 magnitude bits. Multiplying two such numbers can be viewed as the weighted sum of 49 (i.e., 7 × 7) bitwise multiplications. For each bitwise multiplication, if either operand is 0, the result is also 0. This is considered an invalid computation, which can be skipped using sparsity-aware techniques.
\emph{Skipped Calculations} is defined as the ratio of skipped single-bit multiplications to the total 49 single-bit multiplications. 

\label{section:5.3}
\begin{figure}[t]
\centerline{\includegraphics[width=0.5\textwidth]{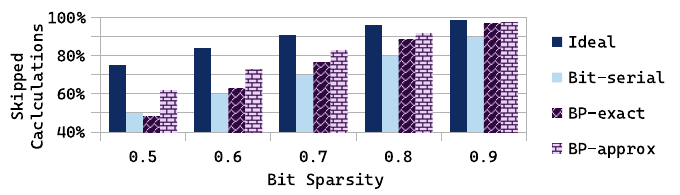}}
\caption{Ratio of skipped 1-bit $\times$ 1-bit calculations for different sparsity-driven approach (range is limited to 40\%-100\%, 
 and higher is better)}
\label{fig:eval_6}
\end{figure}

Fig. \ref{fig:eval_6} presents the experimental results. Here, ``Ideal'' represents a hypothetical scenario where all invalid computations are skipped while retaining all valid ones. However, implementing such an ideal approach incurs prohibitive control logic overhead, negating any potential benefits. Thus, this metric serves only as a reference, demonstrating the upper limit of sparsity exploitation. The ``Bit-serial'' represents a fundamental variant of sparsity-aware bit-serial approaches, where zeros in one operand are entirely skipped. 

The evaluation results indicate that BP-exact surpasses the Bit-serial in terms of bit-level sparsity exploitation for data with over 52\% sparsity. Since 2-bit particles must be entirely zero to be exploited, the proposed approach still falls short of the ideal scenario. For data with 60\%-90\% sparsity, it achieves 74.5\%, 84.0\%, 92.0\%, and 97.7\% of the ``Ideal'', while the bit-serial method only reaches 71.4\%, 76.9\%, 83.3\%, and 90.9\%. 
For the approximate variant, a substantial number of computations (including some valid ones) are skipped, leading to better performance on the Skipped Calculations metric, though at the cost of reduced accuracy. Therefore, it should be applied in scenarios where such trade-off is acceptable.

\subsection{Enhanced Area Efficiency at Comparable Energy Efficiency for the MAC Array}
\label{section:5.4}

\begin{figure}[t]
\centerline{\includegraphics[width=0.52\textwidth]{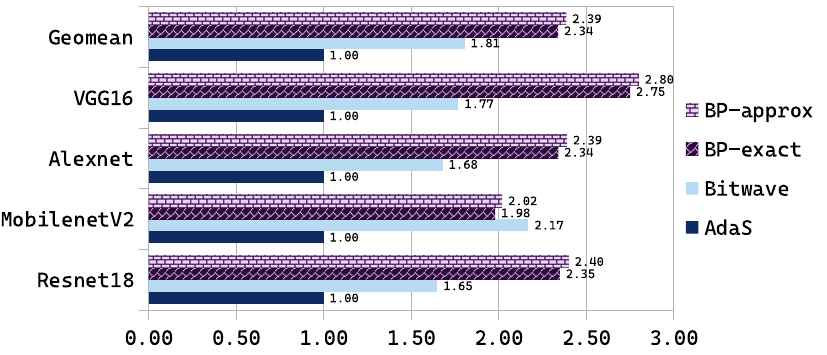}}
\caption{Normalized area efficiency compared to state-of-the-art approaches}
\label{fig:eval_8_1}
\end{figure}

\begin{figure}[t]
\centerline{\includegraphics[width=0.52\textwidth]{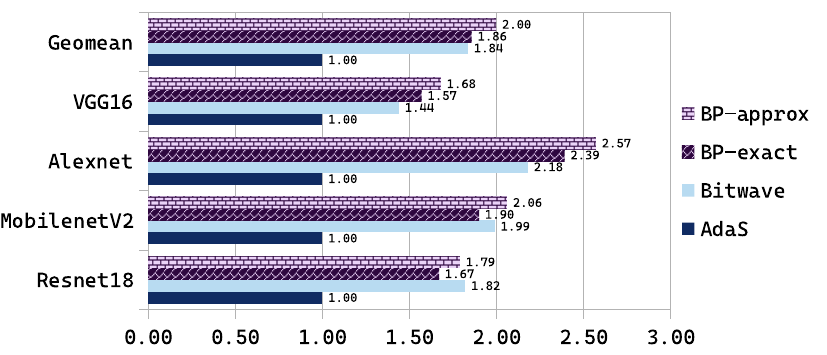}}
\caption{Normalized energy efficiency compared to state-of-the-art approaches}
\label{fig:eval_8_2}
\end{figure}

This section presents a comparison of the proposed MAC array architecture with state-of-the-art bit-sparsity-driven architectures in terms of area efficiency (Fig. \ref{fig:eval_8_1}) and energy efficiency (Fig. \ref{fig:eval_8_2}), with all the data normalized to AdaS. First, BitParticle improves area efficiency by 29.2\% over BitWave while maintaining comparable energy efficiency (geometric mean). The MAC unit of BitParticle, with a significant area advantage over BitWave, is the primary driver of this improvement.
Although BitParticle and BitWave achieve similar energy efficiency, BitWave requires software preprocessing, complicating deployment and making it unsuitable for scenarios with frequent weight updates.
The lower performance of BitParticle on MobileNetV2 is primarily attributed to the network’s extremely low value sparsity, which limits the potential for relieving queue pressure through zero-value filtering. Nevertheless, as most networks exhibit a non-negligible level of value sparsity, this limitation does not undermine the general applicability of the proposed approach.

Second, compared to AdaS, BitParticle achieves significant improvements in both area efficiency and energy efficiency, with improvements of 134\% and 86\%, respectively. 
In addition to the advantages of its MAC units, BitParticle supports multiple dataflows that efficiently adapt to layers of various shapes, whereas AdaS suffers from poor PE utilization on certain shaped layers.
Finally, the approximate variant of BitParticle further improves area and energy efficiency by 2.1\% and 7.5\%, respectively, compared to the exact version.

\section{Conclusions}
This study proposes a novel MAC unit that efficiently exploits dual-factor bit sparsity through particlization-based multiplication. By reducing the number of partial products through grouping and concatenation, the proposed MAC unit achieves superior area and energy efficiency. Moreover, two types of dataflow and a quasi-synchronization approach are proposed to better utilize the MAC unit, achieving improvements of 29.2\% in area efficiency and maintaining comparable energy efficiency over state-of-the-art solutions. 
We hope this approach can inspire further research on more effective exploitation of dual-factor bit sparsity.
\begingroup
\renewcommand{\baselinestretch}{0.9}
\bibliographystyle{IEEEtran} 
\bibliography{references}

\begin{thebibliography}{10}
\providecommand{\url}[1]{#1}
\csname url@samestyle\endcsname
\providecommand{\newblock}{\relax}
\providecommand{\bibinfo}[2]{#2}
\providecommand{\BIBentrySTDinterwordspacing}{\spaceskip=0pt\relax}
\providecommand{\BIBentryALTinterwordstretchfactor}{4}
\providecommand{\BIBentryALTinterwordspacing}{\spaceskip=\fontdimen2\font plus
\BIBentryALTinterwordstretchfactor\fontdimen3\font minus \fontdimen4\font\relax}
\providecommand{\BIBforeignlanguage}[2]{{%
\expandafter\ifx\csname l@#1\endcsname\relax
\typeout{** WARNING: IEEEtran.bst: No hyphenation pattern has been}%
\typeout{** loaded for the language `#1'. Using the pattern for}%
\typeout{** the default language instead.}%
\else
\language=\csname l@#1\endcsname
\fi
#2}}
\providecommand{\BIBdecl}{\relax}
\BIBdecl

\bibitem{quantized_review_1}
T.~Liang, J.~Glossner, L.~Wang, S.~Shi, and X.~Zhang, ``Pruning and quantization for deep neural network acceleration: A survey,'' \emph{Neurocomputing}, vol. 461, pp. 370--403, 2021.

\bibitem{quantized_review_2}
G.~Armeniakos, G.~Zervakis, D.~Soudris, and J.~Henkel, ``Hardware approximate techniques for deep neural network accelerators: A survey,'' \emph{ACM Comput. Surv.}, vol.~55, no.~4, Nov. 2022.

\bibitem{value_sparsity_1_trapezoid}
Y.~Yang, J.~S. Emer, and D.~Sanchez, ``Trapezoid: A versatile accelerator for dense and sparse matrix multiplications,'' in \emph{2024 ACM/IEEE 51st Annual International Symposium on Computer Architecture (ISCA)}, 2024, pp. 931--945.

\bibitem{value_sparsity_2_hetro_gnn}
P.~Chen, P.~Manjunath, S.~Wijeratne, B.~Zhang, and V.~Prasanna, ``Exploiting on-chip heterogeneity of versal architecture for gnn inference acceleration,'' in \emph{2023 33rd International Conference on Field-Programmable Logic and Applications (FPL)}, 2023, pp. 219--227.

\bibitem{value_sparsity_3_scnn}
A.~Parashar, M.~Rhu, A.~Mukkara, A.~Puglielli, R.~Venkatesan, B.~Khailany, J.~Emer, S.~W. Keckler, and W.~J. Dally, ``Scnn: An accelerator for compressed-sparse convolutional neural networks,'' in \emph{2017 ACM/IEEE 44th Annual International Symposium on Computer Architecture (ISCA)}, 2017, pp. 27--40.

\bibitem{bit_sparsity_1_bitwave}
M.~Shi, V.~Jain, A.~Joseph, M.~Meijer, and M.~Verhelst, ``Bitwave: Exploiting column-based bit-level sparsity for deep learning acceleration,'' in \emph{2024 IEEE International Symposium on High-Performance Computer Architecture (HPCA)}, 2024, pp. 732--746.

\bibitem{bit_sparsity_2_pragmatic}
J.~Albericio, A.~Delm\'{a}s, P.~Judd, S.~Sharify, G.~O'Leary, R.~Genov, and A.~Moshovos, ``Bit-pragmatic deep neural network computing,'' ser. MICRO-50 '17.\hskip 1em plus 0.5em minus 0.4em\relax Association for Computing Machinery, 2017, p. 382–394.

\bibitem{bit_sparsity_adas}
X.~Lin, G.~Li, Z.~Liu, Y.~Liu, F.~Zhang, Z.~Song, N.~Jing, and X.~Liang, ``Adas: A fast and energy-efficient cnn accelerator exploiting bit-sparsity,'' in \emph{2023 60th ACM/IEEE Design Automation Conference (DAC)}, 2023, pp. 1--6.

\bibitem{bit_sparsity_3_bitlet}
H.~Lu, L.~Chang, C.~Li, Z.~Zhu, S.~Lu, Y.~Liu, and M.~Zhang, ``Distilling bit-level sparsity parallelism for general purpose deep learning acceleration,'' in \emph{MICRO-54: 54th Annual IEEE/ACM International Symposium on Microarchitecture}, ser. MICRO '21.\hskip 1em plus 0.5em minus 0.4em\relax New York, NY, USA: Association for Computing Machinery, 2021, p. 963–976.

\bibitem{bit_sparsity_4_fusekna}
J.~Yang, Z.~Zhang, Z.~Liu, J.~Zhou, L.~Liu, S.~Wei, and S.~Yin, ``Fusekna: Fused kernel convolution based accelerator for deep neural networks,'' in \emph{2021 IEEE International Symposium on High-Performance Computer Architecture (HPCA)}, 2021, pp. 894--907.

\bibitem{bit_sparsity_ebs}
N.~Jing, Z.~Zhang, Y.~Sun, P.~Liu, L.~Chen, Q.~Wang, and J.~Jiang, ``Exploiting bit sparsity in both activation and weight in neural networks accelerators,'' \emph{Integration}, vol.~88, pp. 400--409, 2023.

\bibitem{bit_sparsity_laconic}
S.~Sharify, A.~D. Lascorz, M.~Mahmoud, M.~Nikolic, K.~Siu, D.~M. Stuart, Z.~Poulos, and A.~Moshovos, ``Laconic deep learning inference acceleration,'' ser. ISCA '19.\hskip 1em plus 0.5em minus 0.4em\relax New York, NY, USA: Association for Computing Machinery, 2019, p. 304–317.

\bibitem{dataflow_systolic}
Kung, ``Why systolic architectures?'' \emph{Computer}, vol.~15, no.~1, pp. 37--46, 1982.

\bibitem{eval_model_resnet}
K.~He, X.~Zhang, S.~Ren, and J.~Sun, ``Deep residual learning for image recognition,'' in \emph{2016 IEEE Conference on Computer Vision and Pattern Recognition (CVPR)}, 2016, pp. 770--778.

\bibitem{eval_dataset_cifar10}
A.~Krizhevsky, ``Learning multiple layers of features from tiny images,'' 2009.

\bibitem{system_loop_tiling}
M.~Wolfe, ``More iteration space tiling,'' in \emph{Supercomputing '89:Proceedings of the 1989 ACM/IEEE Conference on Supercomputing}, 1989, pp. 655--664.

\bibitem{eval_zigzag}
L.~Mei, P.~Houshmand, V.~Jain, S.~Giraldo, and M.~Verhelst, ``Zigzag: Enlarging joint architecture-mapping design space exploration for dnn accelerators,'' \emph{IEEE Transactions on Computers}, vol.~70, no.~8, pp. 1160--1174, 2021.

\bibitem{eval_model_mobilenetv2}
M.~Sandler, A.~Howard, M.~Zhu, A.~Zhmoginov, and L.-C. Chen, ``Mobilenetv2: Inverted residuals and linear bottlenecks,'' in \emph{Proceedings of the IEEE Conference on Computer Vision and Pattern Recognition (CVPR)}, June 2018.

\bibitem{eval_model_alexnet}
A.~Krizhevsky, I.~Sutskever, and G.~E. Hinton, ``Imagenet classification with deep convolutional neural networks,'' \emph{Commun. ACM}, vol.~60, no.~6, p. 84–90, May 2017.

\bibitem{eval_model_VGG}
K.~Simonyan and A.~Zisserman, ``Very deep convolutional networks for large-scale image recognition,'' in \emph{International Conference on Learning Representations}, 2015.

\bibitem{eval_cacti}
N.~P. Jouppi, A.~B. Kahng, N.~Muralimanohar, and V.~Srinivas, ``Cacti-io: Cacti with off-chip power-area-timing models,'' \emph{IEEE Transactions on Very Large Scale Integration (VLSI) Systems}, vol.~23, no.~7, pp. 1254--1267, 2015.

\bibitem{eval_dataset_stl10}
A.~Coates, A.~Ng, and H.~Lee, ``An analysis of single-layer networks in unsupervised feature learning,'' in \emph{Proceedings of the Fourteenth International Conference on Artificial Intelligence and Statistics}, ser. Proceedings of Machine Learning Research, G.~Gordon, D.~Dunson, and M.~Dudík, Eds., vol.~15.\hskip 1em plus 0.5em minus 0.4em\relax Fort Lauderdale, FL, USA: PMLR, 11--13 Apr 2011, pp. 215--223.

\end{thebibliography}
\endgroup
\end{document}